# Early Recognition of Emerging Flu Strain Clusters

A.   Li[(1)], J. C. Phillips[(2)], and M. W. Deem[(1,3)]

(1)  Departments of Bioengineering and (3) Physics & Astronomy, Rice University,Houston, TX 77005
(2) Dept. of Physics and Astronomy, Rutgers University, Piscataway, N. J., 08854

Abstract

Minimizing time delays in manufacturing vaccines appropriate to rapidly mutating viruses is the key step for improving vaccine effectiveness.  The vaccine for the H3N2 flu type has failed for the last two years (~ 15% effective). Here we summarize the state of the predictive art and report the most current results for H3N2 flu vaccine design.  Using a 2006 model of dimensional reduction of viral mutational complexity, we show that this model can reduce vaccine time delays by a year or more in some cases.

Viruses are hypermutated proteins, with nearly one mutation per replication [1].    This rapid mutation allows selection for viruses that succeed in evading antibody-mediated immunity.  In a computational view the most successful strains tend to form clusters that are visible in a reduced, two-dimensional sequence space.  This cluster of strains remains until sufficiently recognized by antibodies in the human population [2], at which point selection leads to a new antibody-evading strain, which initiates a new cluster.   This process is a dramatic example of punctuated evolution [3-5], and it is modelled with cluster-specific immunity [6].  There are several types of flu viruses, and the current vaccine cocktail treats the most common three.  The oldest type A/H1N1 was responsible for both the deadly 1918 and the comparatively harmless 2009 "swine flu" pandemics.  The A/H1N1 type strains have stabilized, are milder, and are more effectively vaccinated against.  More morbidity and mortality are now associated with rapidly mutating A/H3N2 strains, which are discussed here.

There are two major problems facing the vaccination community.  The first is the lengthy temporal delay between the arrival of new evasive viral strains, their reported sequencing, and determination of suitability of new strains as vaccine targets.  The second is the effect of mutations in the currently circulating strains on the effectiveness of vaccines based on older strains, often including another lengthy delay due to collection of ferret animal model data. Traditionally these two problems have most often been treated separately, while a unified treatment could result in a dramatic acceleration of vaccine design.

Early strain clustering efforts utilized both indirect measurements of viral activity, for instance, by measuring ferret responses, as well as molecular evolution at the amino acid sequential level [7,8]. In the latter method, sequences are unambiguous, and the historical sequence data base has grown to be much larger and more timely than the still sparse and lagging-by-years ferret viral activity database [9,10]. As a result, since 2006 both improved clustering and better dominant strain predictive results have been obtained bioinformatically for H3N2 using sequences alone. Until 2010, annual H3N2 differences were small, and the two methods were almost equally successful (8 out of 10 years for the mixed ferret-sequence method, improved to 9 out of 10 for the pure sequence method) [11,12].

Sequential substitutional displacements can be measured in different ways. One measure of distance between any two sequences is the number of mutations [9]. The most substitutionally active part of flu viruses is the ~ 100 amino acid head region 115-215 of the hemagglutinin protein, HA1, which contains 329 amino acids. A more refined measure of difference between strains begins by identifying the five HA1 epitopic subsets (A-E), containing the most frequently mutated amino acids in a given year, with the A and B epitope patches concentrated at the top of the head region, near the sialic acid receptor [7,8].

The best correlation with vaccine effectiveness to date defines distances between sequences in terms of the numbers of mutations at epitopic sites, normalized by the size of the epitope, and identifies the epitope that is dominating the difference between the current circulating strain and the lagging vaccine strain [8]. The fractional number of all mutations of this dominant epitope, $p_{epitope}$, gives the best correlation with vaccine efficacy [7,9], with $R^2$ ~ 0.8 over the period 1971-2015, which is much better than one obtains from ferret data, or mixed ferret- sequence data, $R^2$ ~ 0.5 [8].

An accurate and rapid way to quantify strain drift [9,10], and at the same time recognize formation of clusters of new strains, is to compress dimensionally the interstrain distances in the full amino acid space (for A/H3N2, 329 dimensions) to the two eigenvectors best capable of minimizing the interstrain distances over a time frame of about three years [7,11]. Dimensional compression of sequences is unbiased, and is compatible with epitopic analysis [11,12]. Here the compression from 329 dimensions to only two dimensions can be regarded as an extreme case, which can introduce both common and distinctive features as new data are added. This point is discussed in Figs. 1 and 2, showing strain cluster evolution over 3 months in 2014-2015.

The most obvious feature of Fig. 1 (2012- Fall 2014) is the large separation between Texas 2012 and the central 2013-2014 cluster, with $p_{epitope}$ = 0.19 (derived from the B epitope), which explains why the 2013 and 2014 vaccines based on Texas 2012 have failed [13]. The strain selected by the World Health Organization as the target for the 2015 vaccine is Switzerland 2013. This was a novel cluster in 2014, but there is another novel cluster in 2014-2015, centered near Nebraska 2014. The $p_{epitope}$ separating these two different 2015 candidate clusters (derived from the A epitope) is 0.16, only slightly smaller than vaccine target and dominant strain $p_{epitope}$ = 0.19 in failed vaccine years2013-2014.

Next we look carefully at the Nebraska 2014 cluster, and compare with Switzerland 2013. Which better represents the dominant circulating strain for 2015? In Fig. 2 we show what happens when the late 2014 strains reported in early 2015 are added to Fig. 1. The axes are rotated, and Switz moves closer to centered earlier strains, while >90% of the new strains are added to the well-separated Nebraska 2014 cluster. This is still far from the vaccine target strain Switz 2013 cluster, with a separation value of $p_{epitope}$ = 0.16 large enough to predict vaccine failure again in 2015-2016 [13].

It is tempting to suppose that perhaps better predictability could be achieved by including deleterious (mostly C-E epitopes) mutations outside the A-B receptor region phylogenetically [14]. In this way, and using observed sequence data to estimate clade "fitness," some encouraging H3N2 results (inset to their Fig. 3) were obtained. These results stopped short of reaching a correlation with vaccine effectiveness similar to that already reported earlier [8] for A/H3N2 using all epitopes. There are several toy models of network "fitness" [15], but they all involve only short-range connectivity, and no long-range forces. Long-range forces can be modelled using hydropathic profiles, but as we shall soon see, they are not needed here for predicting H3N2 vaccine targets [16-19]. Note the two striking nearly linear regions associated with different years in Fig. 2. These were already present in Fig. 1, but the lines have straightened and narrowed in Fig. 2. This novel feature deserves more study. It is unlikely to be accidental: it could represent the effect of long-range interactions on the dominant epitope, which changed from B to A from 2012-2013 to 2014 [13].

Why is the dominant epitope model so effective?  A toy model describes the A-B epitope dominance switching as vector "canalization" [20], which could be generalized to identify a single antigenic axis for each dominant epitope.    Markov analysis explored this question at length for the historical H3N2 switching between A and B dominant epitopes (1971-2004), and two results emerged [12]:  (1) The A and B epitopes both combine hydrophobic and charged amino acid features, but with inverted conserved hydrophobic structures (epitopic corners A, and center B); and  (2) There was a positive selective pressure for increasing charged amino acids in the immunodominant epitopes, which decrease absolute values of binding energies of epitopes to antibodies, by increasing the epitope's affinity for water.  Further examination of A and B epitopes, 2004-2014 [13], reveals qualitative differences in A-B switching before and after 2000, which will be discussed elsewhere.

In conclusion, rapid and reliable identification of emergent strain clusters is the key to developing an effective flu vaccine.  Traditional measurements of viral antigenicity or phylogenetic theories, or their combination, omit highly informative structural data used to define epitopes [22-24], and have had only limited success ($R^2$ ~ 0.5).  Dimensional reduction in three- year time-frame sliding sequence space, supported by identifying dominant epitopes [8,11-13], has proved to be more effective for H3N2 ($R^2$ ~ 0.8).   The stakes here are high: a one-year delay in finding new H3N2 clusters has an equivalent economic value > $10 billion (a conservative estimate, considering that H3N2 morbidity is several times higher than for H1N1 [25]).   The failure of the H3N2 2014-2015 vaccine in America was responsible for 6 million additional flu cases lasting several weeks, disrupted many hospitals, and undermined public confidence in vaccinations.

Because of record late Nov. - Dec. temperature highs (~ 10°C above normal) in most of America, 2015-2016 American flu infections, reported online weekly at CDC FLUVIEW, are ambiguous at this writing. We predict that when winter temperatures return to normal, 2015 flu epidemic severity will be less than 2014, with $p_{epitope}$ = 0.19, but more than 2013, with $p_{epitope}$ = 0.14, as predicted by the A/Switzerland-Nebraska difference with $p_{epitope}$ = 0.16 [13].

Key words: vaccine, amino acid sequence, search algorithms

# References

1. Drake, JW; Holland, JJ (1999) Mutation rates among RNA viruses. Proc. Natl. Acad. Sci. USA **96**, 13910-13913.

2. Bianconi, G; Fichera, D; Franz, S; Peliti, L (2011) Modeling microevolution in a changing environment: the evolving quasispecies and the diluted chamption process. J. Stat. Mech: Theory Exper. **11** P08022.

3. Bak, P; Sneppen, K (1993) Punctuated equilibrium and criticality in a simple model of evolution. Phys. Rev. Lett. **71**, 4083-4086.

4. Elena, SF; Cooper, VS; Lenski, RE (1996) Punctuated evolution caused by selection of rare beneficial mutations. Science **272**, 1802-1804.

5. Moret, MA; Pereira, HBB; Monteiro, SL; Galeao, AC (2012) Evolution of species from Darwin theory: A simple model. Phys. A-Stat. Mech. Appl. **391**, 2803-2806.

6. Koelle, K; Cobey, S; Grenfell, B; et al. (2006) Epochal evolution shapes the phylodynamics of interpandemic influenza A (H3N2) in humans. Science **314**, 1898-1903.

7. Smith, DJ; Lapedes, AS; de Jong, JC; et al. (2004) Mapping the antigenic and genetic evolution of influenza virus. Science **305**, 371-376.

8. Gupta, V; Earl, DJ; Deem, MW (2006) Quantifying influenza vaccine efficacy and antigenic distance. Vaccine **24**, 3881-3888.

9. Skowronski, D. M.; Janjua, N.Z.; De Serres, Gaston; et al. (2014) Low 2012-13 Influenza vaccine effectiveness associated with mutation in the egg-adapted H3N2 vaccine strain not antigenic drift in circulating viruses. PLoS ONE **9**, E92153.

10. Xie, Hang; Wan, Xiu-Feng; Ye, Zhiping; et al. (2015) H3N2 Mismatch of 2014-15 Northern Hemisphere Influenza Vaccines and Head-to-head Comparison between Human and Ferret Antisera derived Antigenic Maps. Scien. Rep. 5, 15279.

11. He, J; Deem, MW (2010) Low-dimensional clustering detects incipient dominant influenza strain clusters. Prot. Eng. Design, Selec. **23**, 935-946.

12. Pan, K; Long, J; Sun, H; Tobin, GJ; Nara, PL; Deem, MW (2011) Selective pressure to increase charge in immunodominant epitopes of the H3 hemagglutinin influenza protein. J. Mol. Evol. **72**, 90-103.

13. Li, X; Deem, MW (2015), Influenza evolution and vaccine effectiveness in the 2014/2015 Season. arXiv 1510.00488.

14. Luksza, M; Laessig, M (2014) A predictive fitness model for influenza. Nature **507**, 57–61.

15. Caldarelli, G; Capocci, A; De Los Rios, P; et al. (2002) Scale-free networks from varying vertex intrinsic fitness. Phys. Rev. Lett. **89**, 258702.

16. Moret, MA; Zebende GF (2007) Amino acid hydrophobicity and accessible surface area. Phys. Rev. E **75**, 011920.

17. Phillips, JC (2014) Fractals and self-organized criticality in proteins. Phys A **415**, 440-448.

18. Phillips, JC (2014) Punctuated evolution of influenza virus neuraminidase (A/H1N1) under opposing migration and vaccination pressures. BioMed Research International **2014**, 907381.

19. Phillips, JC (2015) Similarity is not enough: Tipping points of Ebola Zaire mortalities. Phys. A **427,** 277-281.

20. Bedford, T; Rambaut, A; Pascual, M (2012) Canalization of the evolutionary trajectory of the human influenza virus. BMC Biol. **10,** 38.

21. Lee, Eva K.; Yuan, Fan; Pietz, Ferdinand H.; et al. (2015) Vaccine prioritization for effective pandemic response. Interfaces **45**, 425-443.

22. Hopp TP, Woods KR (1981) Prediction of protein antigenic determinants from amino acid sequences. Proc. Nat. Acad. Sci. (USA) **78**, 3824-3828.

23. Munoz, E; Deem, MW (2005) Epitope analysis for influenza vaccine design. Prot. Eng. Des. Selec. **21**, 311-317.

24. Correia, Bruno; Bates, John; Loomis, Rebecca; et al. (2014) Proof of principle for epitope-focused vaccine design. Nature **507**, 201-206.

25. Pop-Vicas, Aurora; Rahman, Momotazur; Gozalo, Pedro L.; et al. (2015) Estimating the effect of influenza vaccination on nursing home residents' morbidity and mortality. J. Am. Ger. Soc. **63**, 1798-1804.

# Figure Captions

Fig. 1.   Dimensional reduction of all HA1 sequences deposited in Genbank in the time frame from 2012 through 12/2014 [13].  As the color coding shows, the Nebraska 2014 cluster began to appear as a few strains by Dec 2013, and contained > 10 strains by Mar 2014.  The Cal 2014 cluster also contains the WHO vaccine target strain Switz 2013.

Fig. 2.   Most of the 877 late 2014 strains reported in early 2015 are concentrated in the Nebraska 2014 cluster, which is far separated from the Switz 2013 cluster.  These figures are best viewed online at high magnification.  In addition to the two clusters there are also two linear "canals" associated with 2012-early 2013, and late 2014.  The largest difference between Figs. 1 and 2 is merely rotation of the coordinates.  The origins are the center of masses, which are little changed.  For clarity the graphical resolution of 4754 points (877 added to Fig. 1) is limited.

Fig. 1

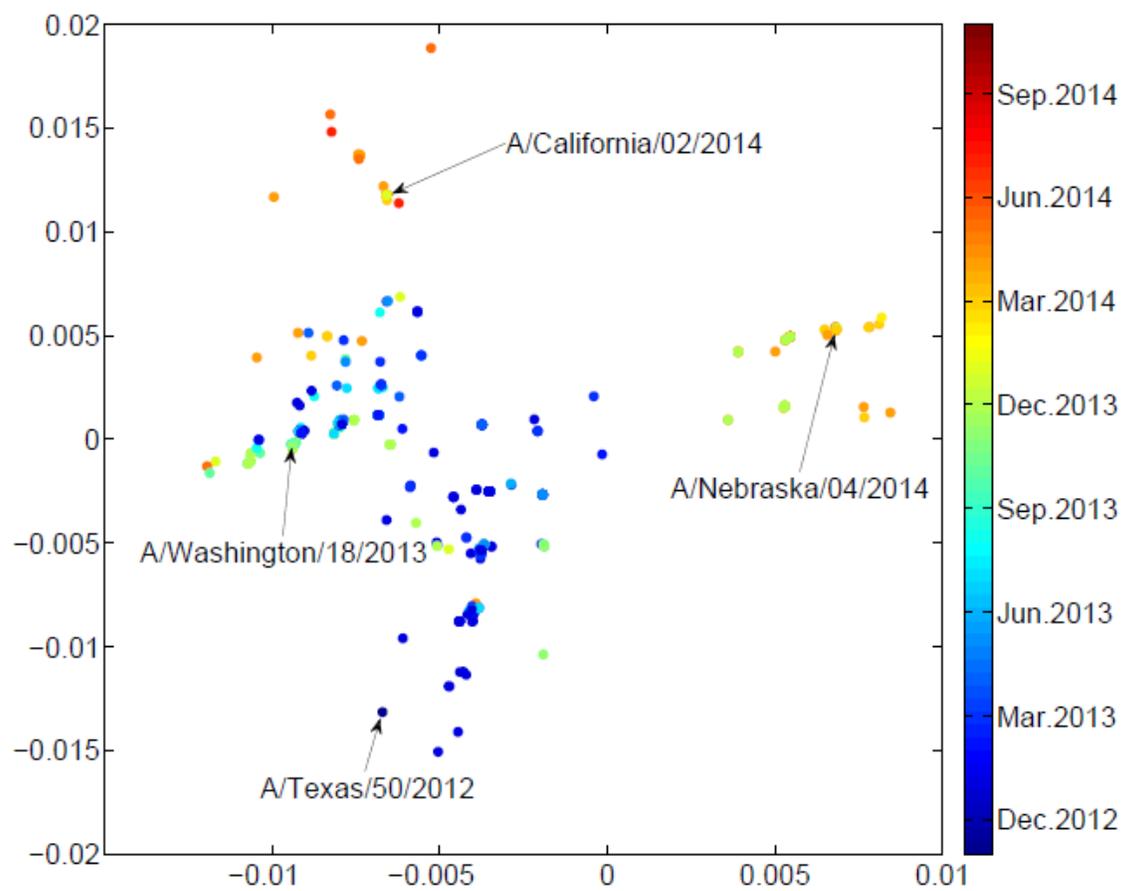

Fig. 2

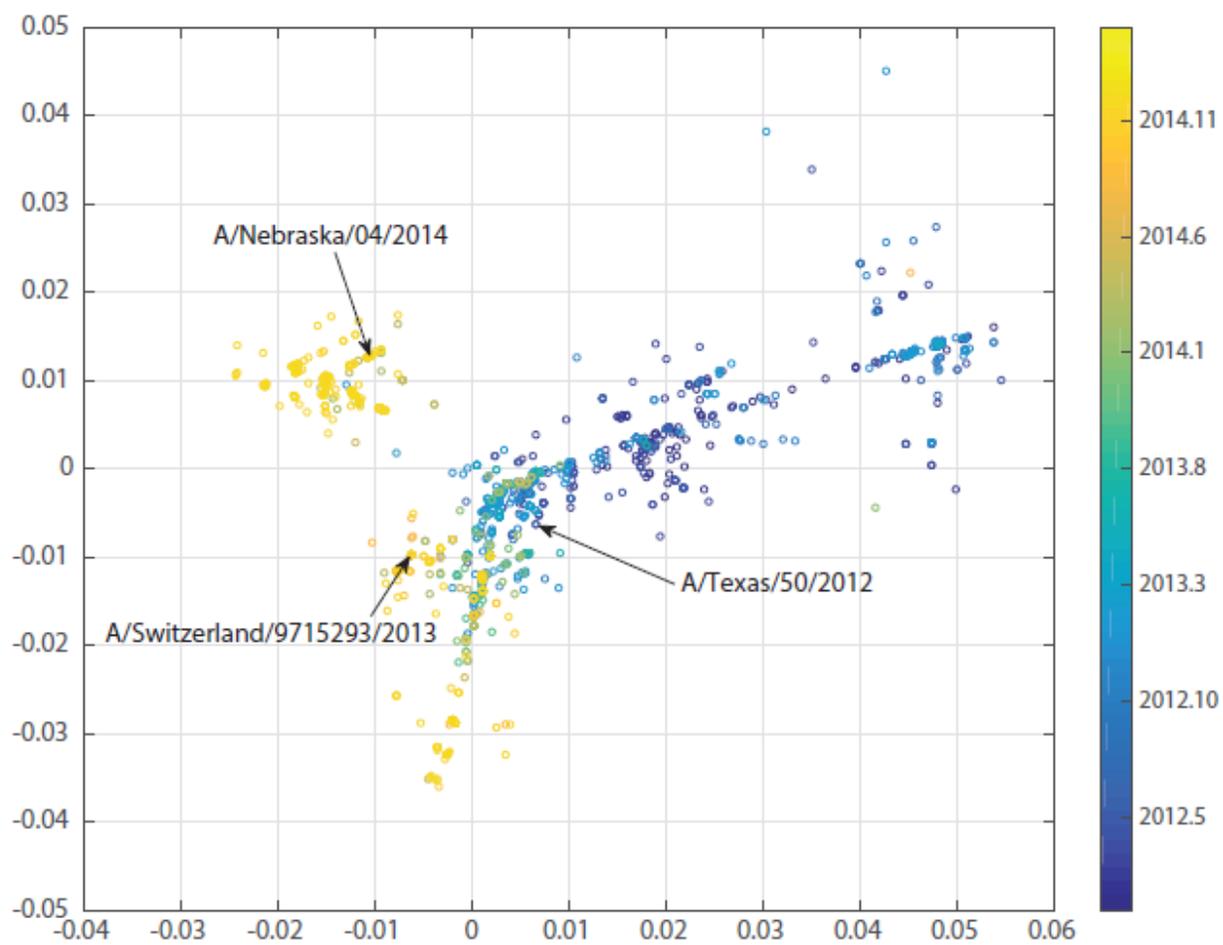